\documentclass{article}
\usepackage[utf8]{inputenc}
\usepackage{amsmath}
\usepackage{amsthm, amssymb}    
\usepackage{graphicx,tikz}
\usepackage{times, verbatim}    

\newtheorem{theorem}{Theorem}

\usepackage{mathtools}
\usepackage{hyperref}
\usepackage{xcolor}
\usepackage[sorting=none,style=numeric]{biblatex}
\usepackage[hcentering, bindingoffset = 10mm, right = 20 mm, left = 7 mm, top=20mm, bottom = 20 mm]{geometry}

\usetikzlibrary{math} 

\tikzmath{\r = .65; \rr = 0.17; \alp = 0.1; \x1 = 0; \y1 =0; \x2 = 2; \y2 =3.4641; \x3 = 4; \y3 =0; 
\xv = \x1 + \alp*\x2; \yv = \y1 + \alp*\y2;
\xvp = \x1 + \alp*\x3; \yvp = \y1 + \y3 - \alp*\y3; \xwp = \x1 + \x3 - \alp*\x3; \ywp = \y1 + \y3 -\alp*\y3; \xu = \x1 + \x2 - \alp*\x2; \yu = \y1 + \y2 - \alp*\y2; \xup = \x2 + \alp*\x2; \yup = \y2-\alp*\y2; \xw = \x2 + \x2- \alp*\x2; \yw = \y2-\y2 +\alp*\y2;}

\usepackage{hyperref}
\def \tr{{\mbox{tr~}}}

\def \be{\begin{equation}}
\def \ee{\end{equation}}
\def \bea{\begin{eqnarray}}
\def \eea{\end{eqnarray}}

\newcommand{\cpn}{\mathbb{C}P^n}
\newcommand{\im}{\operatorname{Im}}

\title{Geometry of degenerate quantum states, configurations of $m$-planes and invariants on complex Grassmannians}

\begin{document}

\author{Alexander Avdoshkin\\
Department of Physics, Massachusetts Institute of Technology, Cambridge, MA 02139, USA\\
Department of Physics, University of California, Berkeley, CA 94720, USA}

\maketitle

\begin{abstract}
{Understanding the geometric information contained in quantum states is valuable in various branches of physics, particularly in solid-state physics when Bloch states play a crucial role. While the Fubini-Study metric and Berry curvature form offer comprehensive descriptions of non-degenerate quantum states, a similar description for degenerate states did not exist. In this work, we fill this gap by showing how to reduce the geometry of degenerate states to the non-abelian (Wilczek-Zee) connection $A$ and a previously unexplored matrix-valued metric tensor $G$. Mathematically, this problem is equivalent to finding the $U(N)$ invariants of a configuration of subspaces in $\mathbb{C}^n$. For two subspaces, the configuration was known to be described by a set of $m$ principal angles that generalize the notion of quantum distance. For more subspaces, we find $3 m^2 - 3 m + 1$ additional independent invariants associated with each triple of subspaces. Some of them generalize the Berry-Pancharatnam phase, and some do not have analogues for 1-dimensional subspaces. We also develop a procedure for calculating these invariants as integrals of $A$ and $G$ over geodesics on the Grassmannain manifold. Finally, we briefly discuss possible application of these results to quantum state preparation and $PT$-symmetric band structures.

}
\end{abstract}

\section{Introduction}

In modern theoretical physics one often encounters a collection of quantum states, or a quantum state depending on a parameter. Adiabatic evolution\cite{berry1984quantal}, band structures in crystals\cite{RevModPhys.82.1959} and molecular electronic wave-functions in the Born-Oppenheimer approximation \cite{born2000quantum} are some prominent examples. When no extra structure (e.g. operators, locality) is assumed, the observables are conveniently expressed in terms of geometric objects that describe the quantum space of states.

Fundamentally, geometric structures arise when one studies a set of states up to unitary transformations (basis changes) of the Hilbert space. Geometry here is understood in the sense of the Erlangen program \cite{erlangen}, i.e. we study invariants on some space under the action of its symmetry group. In our context, the space consists of the normalized elements of the Hilbert space modulo the phase \if state space is the complex projective space or the complex Grassmannian depending on whether we are considering non-degenerate or degenerate states \fi and the symmetry group is the unitary group that preserves the inner product structure of the Hilbert space.

The geometric perspective has recently gained popularity as a tool for calculating physical properties of crystalline materials. In the past, the main focus has been on the topology of the wavefunctions and its connection with the presence of ungappable edge modes in topological insulators \cite{top-insulators-review}. Currently, there is an effort to understand the effect of electronic wavefunctions on material properties beyond topology. One example is flat band materials, where the band dispersion vanishes, and these effects can be particularly strong. In these materials the conditions for formation of correlated states and their properties were described in terms of geometric structures \cite{roy2014band, flat-band-superconductor}. Another example is non-linear optical responses, which are highly sensitive to wavefunction shapes (i.e. the real space structure implied by the Bloch states) and require a thorough understanding of quantum state geometry \cite{ahn-1, ahn-2, AvdMoore}. Finally, the objects arising in the modern theory of polarization allow a very compact representation via geometric objects \cite{souza2000polarization, avdoshkin2022}.

The geometric approach essentially consists in reducing the observables to the invariants of the collection of states under unitary basis changes. In order for the approach to be useful, one needs to have a parametrization of the independent invariants and an efficient way to compute them. In the case of non-degenerate states, the state space is the complex projective space $\cpn$, known as the “ray space” in physics (the Bloch sphere for a two-level system). The complete set of invariants is given by quantum distances computed for every pair of states and Pancharatnam phases \cite{pancharatnam1956generalized, cantoni1992three} computed for every triple of states \cite{\iffalse bengtsson_zyczkowski_2006, \fi avdoshkin2022}. Both these objects can be computed as integrals of the Fubini-Study metric and a symplectic (Berry curvature) form. We review this case in Sec. \ref{sec:one_d}.

It is also not uncommon in physics for the quantum state of interest to become degenerate in energy with other states. In band structures, level crossings are generic in 3D and in PT-symmetric materials Kramer's degeneracy leads to bands doubly degenerate for all quasimomenta \cite{kramers1930theorie}. Similarly, the configuration of non-abelian anyons only specifies the subspace and not the exact quantum state \cite{frohlich1988statistics}. However, neither the invariants characterizing a collection of degenerate states, i.e. $m$-dimensional subspaces of the Hilbert space, nor their representation in terms of local tensors was known, thus hindering the application of the geometric approach. The goal of this paper is to develop such an approach.

We begin by identifying all invariants characterizing a collection of $m$-dimensional subspaces of an $n$-dimensional Hilbert space,  Sec. \ref{sec:invariants}. We reproduce the result of Jordan \cite{jordan1875essai} that for a pair of $m$-planes, when $n$ is sufficiently large, there are $m$ independent invariants, known as principal angles, that generalize the notion of quantum distance. For three $m$-planes, we find $3 m^2 - 3 m + 1$ additional independent objects that can be represented as three unitary matrices, that we call principal unitaries, acting on each of the planes in a particular basis. We are not aware of any discussion of these objects in existing literature. The configuration of $n>3$ subspaces is fully described by finding the invariants for all pairs and triples of subspaces. They are also independent modulo simple relations that we describe. 

It is known that, for degenerate states, the Berry connection, denoted by $A$, becomes non-abelian \cite{PhysRevLett.52.2111} and we will show that the other ingredient needed to calculate the invariants described in the previous paragraph is a matrix-valued version of the Fubini-Study metric $G$. As we show in Sec. \ref{sec:tensors}, together these two objects give rise to all $U(n)$-invariant and co-variant objects on the Grassmannian. We are similarly not aware of any literature that discusses the geometry associated with matrix-valued metric tensors. We describe the geometry coming from $G$ in the context of the Grassmannian manifold, but our results will not apply to an arbitrary fiber bundle. We leave this question for future study.

We find that, as in the one-dimensional case, going from local tensor to global objects can be conveniently done via geodesics. Remarkably, despite multiple notions of distance on the Grassmannian, there is still a unique way of defining geodesics. In particular, using the matrix valued tensor $G$, one can define $m$ functionally-independent Finsler metrics by taking traces of powers of $G$ and symmetrizing over the tangent indices. They share the same geodesics, but the lengths they assign to the geodesics are different. The $m$-different invariants describing a pair of states come about as the integrals of the Finsler metrics. In the last part of Sec. \ref{sec:tensors}, we deal with the computation of 3 subspace invariants, the most subtle part of the construction. \if Using the true invariants, found in Sec. \ref{sec:invariants},\fi We find that most of them are determined by the $G$-tensor alone, and only the objects that generalize the Berry-Pancharatnam \cite{pancharatnam1956generalized, berry1984quantal, cantoni1992three} phase require the use of $A$. 

In the next section, we will formally define the problem and introduce the necessary mathematical structures.

\section{Defining the problem}

\subsection{One-dimensional subspaces} \label{sec:one_d}
Consider an element of the Hilbert space $|\psi\rangle \in \mathbb{C}^{n+1}$. In quantum mechanics, one assumes it to be normalized: $\langle \psi | \psi \rangle = 1$ and, furthermore, states that only differ by an overall phase are physically equivalent: $ |\psi \rangle \sim e^{i \phi} | \psi \rangle$. The normalization condition constrains the states to lie on a sphere $S^{2n+1} = \mathbb{C}^{n+1}/R_{+}$, where $R_+$ is the set of positive real numbers. The identification over the phase choice further leads to factorization by $U(1)$ leaving us with the complex projective space $\cpn = S^{2n+1}/U(1)$, the space of all $1$-dimensional subspaces. $\cpn$ is a complex manifold of dimension $n$ that is also commonly represented as a coset
\bea
\cpn = \frac{U(n+1)}{U(n)\times U(1)}.
\eea
$\cpn$ is  naturally endowed with a linear bundle structure: the fiber over every point is the one-dimensional subspace that this point corresponds to.

Geometric, or invariant, objects that one can define on $\cpn$ have their origin in the inner product structure on the original $\mathbb{C}^{n+1}$. Given two quantum states represented by normalized vectors $u,v \in \mathbb{C}^{n+1}$ we define the quantum distance $\theta$ via the absolute value of the inner product, according to
\bea \label{eq:dist}
|\langle v | u \rangle| = \cos \theta(u,v),~ \theta \in [0,\pi/2].
\eea
$\theta(u,v)$ is invariant under both phase changes of $u$ and $v$ and the action of the unitary group applied to both vectors. It is also the only such invariant that we can construct for two vectors, in the sense that every other invariant must be a function of $\theta(u,v)$.

Let us now construct the invariants for three states $u,v,w \in \mathbb{C}^{n+1}$. Naturally, there are three pair-wise quantum distances $\theta(u,v), \theta(u,w)$ and $\theta(v,w)$. They are functionally independent similarly to the three legs of a triangle on a real plane. However, on $\cpn$, there is an additional fourth invariant, known as the Pancharatnam phase \cite{pancharatnam1956generalized, cantoni1992three}
\bea  \label{eq:phase}
\varphi(v,u,w) = \arg \langle v | u \rangle \langle u | w \rangle \langle w | v \rangle.
\eea
Together Eq. \eqref{eq:phase} and the three $\theta$'s exhaust all invariants for three subspaces. Furthermore, it was shown for any collection of quantum states $|u_i\rangle$, Eq. \eqref{eq:dist} computed for all pairs of states and Eq. \eqref{eq:phase} computed for all triples of states yield all independent invariants. This is formalized in the following \cite{avdoshkin2022}

\begin{theorem} \label{th:gps_old}
For any two sets of $\{|u_i\rangle\}$ and $\{|v_i\rangle\}$ in $\mathbb{C}^{n+1}$ there exists $U \in U(n+1)$ such that $U |u_i\rangle = |v_i\rangle$ (up to phases), if and only if all two point distances for the first set are the same as for the second set and, similarly, for the three-point phases.\end{theorem}

In this way, the $\theta$'s and $\phi$'s are invariant global objects characterizing collection of ray in $\mathcal{C}^{n+1}$. Now we wish to develop a more compact description of the invariants via local objects. We will define two invariant tensors and show that both Eqs. (\ref{eq:dist}, \ref{eq:phase}) reduce to them. To this end, we introduce a coordinate on $\cpn$ that we denote as $x$ (sometimes we will write $x^{\mu}$). We denote as $|u(x) \rangle$ some normalized state that belongs to the subspace that $x$ represents. All possible choices of $|u(x) \rangle$ differ from each other by a phase and we will refer to a different phase choice $|\tilde{u}(x) \rangle = e^{i \phi(x)} |u(x) \rangle$ as a gauge transformation. The first tensor we need is called the Fubini-Study metric $g_{\alpha\beta}(x)$ and comes from the Taylor expansion of the quantum distance:
\bea
|\langle u(x) | u(x+\delta x) \rangle| = 1 - g_{\alpha\beta}(x) \delta x^{\alpha} \delta x^{\beta} + O(\delta x^3).
\eea
The metric $g$ is gauge-invariant, because it is constructed from the gauge-invariant quantum distance.

The second tensor $\omega_{\alpha \beta}$ comes about as the  infinitesimal version of Eq. \eqref{eq:phase}:
\bea
\varphi(u(x), u(x + \delta x_1), u(x + \delta x_2)) = \omega_{\alpha \beta}(x) \delta x_1^{\alpha} \delta x_2^{\beta} + O(\delta x^3).
\eea
It is a  closed (symplectic) two-form known as the Berry curvature in physics. The objects $g_{\mu\nu}$ and $\omega_{\mu \nu}$ are not fully independent because they obey certain relations, e.g. these coming from the Kähler structure, but we will not discuss them here. At the same time, neither of them can be fully reconstructed from the other, and for that reason, we need to use both of them. As mentioned, $g$ and $\omega$ are invariant under the action of $U(n+1)$ and, vise versa, the action of $U(n+1)$ on $\cpn$ can be recovered by considering all diffeomorphisms of $\cpn$ that preserve $g$ and $\omega$ simultaneously. Because of that, one can say that $g$ and $\omega$ define the geometry of $\cpn$.

We can prove the latter fact explicitly: both Eq. \ref{eq:dist} and Eq. \ref{eq:phase} for any set of points can be computed using $g$ and $\omega$ \cite{avdoshkin2022}:

\begin{theorem}
The quantum distance between two states $u$ and $v$ is given by the length of the geodesic $\gamma(u,v)$ with respect to $g$:
\bea \label{eq:int_for_theta}
\theta(u,v) = \int_{\gamma(u,v)} \sqrt{g_{\alpha\beta} dx_\alpha dx_\beta}.
\eea
The 3-pt phase is given by the flux of the 2-form $\omega$ through the geodesic triangle $\gamma_{u,v,w}$ constructed over the three points:
\bea \label{eq:int_for_phi}
\varphi(u,v,w) = -\int_{\gamma_{u,v,w}} \omega,
\eea
where the surface of $\gamma_{u,v,w}$ can be chosen arbitrarily since $\omega$ is closed.
\end{theorem}

Note that Eq. \eqref{eq:int_for_theta} requires only knowledge of $g$, but Eq. \eqref{eq:int_for_phi} requires the knowledge of both $g$ and $\omega$, because $g$ is needed to construct $\gamma_{u,v,w}$. 

To summarize, we described the complete sets of both global and local invariants that characterize quantum states and showed how they are related to each other. This is what we want to reproduce for degenerate states.

\subsection{m-dimensional subspaces}

Now consider an $m$-dimensional subspace $V = \text{span}\{|\psi_i\rangle\}_{i = 1, \cdots, m}$ of $\mathbb{C}^n$. The set of all $m$-dimensional subspaces of an $n$ dimensional space $\mathbb{C}^n$ is known as the Grassmannian $Gr_{m,n}$ \cite{shafarevich2013basic}. For $m=1$, we recover the complex projective space $Gr_{1,n} = \mathbb{C}P^n$. The Grassmannian similarly admits a coset representation 
\bea \label{eq:grasm_def}
Gr_{m,n} = \frac{U(n)}{U(m) \times U(n-m)}
\eea
and has complex dimension $m (n-m)$. As for $\mathbb{C}P^n$, there is a tautological bundle structure on $Gr_{m,n}$ with the fiber over a point being the corresponding subspace \cite{dubrovin2012modern}. 

 As before for $\mathbb{C}P^{n-1}$, there are no invariants associated with a single subspace (except for its dimension); equivalently, the action of $U(n)$ on $Gr_{m,n}$ is transitive. First, we would like to find an analogue of Theorem \ref{th:gps_old}, i.e. find a set of invariants that uniquely determines the configuration of two or more $m$-dimensional subspaces. 
 \newline
 
\textbf{Question 1:}
{\textit{Given a collection of $m$-dimensional subspaces $V_i$, $i = 1, \cdots, l$ what is the set of global quantities that determine the configuration uniquely to up a unitary transformation? We also want these quantities to be as independent from each other as possible.}}
\newline

In \cite{GALLAGHER1977157}, it was shown that the knowledge of all expressions of the form $\tr[P_{i_1} \cdots P_{i_k}]$, $1 \leq i_1,\cdots,i_k \leq l$, where $P_i$ are the projectors on the corresponding subspaces (we discuss them in detail in the next subsection) and $k$ is an integer, fully determines the configuration. We would like to have a more concise description, since this set of traces is infinite and there are, obviously, multiple relations between them. We will see that it is sufficient to describe the invariants of three subspaces and the invariants of a larger number of subspaces will reduce to the invariants of all pairs and triples. We carry out this analysis in Sec. \ref{sec:invariants}.

The second question is finding all independent invariant tensors on $Gr_{m,n}$.
 \newline
 
\textbf{Question 2:}
{\textit{What are the invariant (or co-variant) local objects on $Gr_{m,n}$? }}
\newline

Roughly, speaking these objects will be the non-abelian generalization of the quantum metric and Berry connection. The non-abelian connection, which defines a co-variant derivative, is commonly discussed in theoretical physics, but the non-Abelian version of a metric is more poorly understood and we discuss its properties in more detail. This is the subject of Sec. \ref{sec:tensors}. 

Finally, we would like to understand how the invariants of a collection of subspaces arise from the local objects.
 \newline
 
\textbf{Question 3:}
{\textit{How can one compute the global invariants from Question 1 using the objects from Question 2? }}
\newline

Geodesics will again be the most essential ingredient of that construction, but due to the large number of new invariants, there are a lot of new features. In  particular, we will see that most new invariants arise from the new metric tensor alone. This is discussed in Sec. \ref{sec:tensors}.

Before deriving our main results, we first introduce the projector formalism that we find to be most convenient for working with subspaces algebraically.

\subsection{Projector formalism}

We consider the $m=1$ case (non-degenerate states) first. Given a normalized state $| \psi \rangle$, the corresponding projector is constructed as $P = | \psi \rangle \langle \psi |$. There is a one-to-one correspondence between the element of $\cpn$ and orthogonal rank-1 projectors, i.e. operators that satisfy: $\text{rank}~ P = 1$, $P^2 = P$ and $P^{\dagger} = P$. In other words, projectors are insensitive to the arbitrary phase choice of wavefunction, but contain all other information about them. This is the main advantage of this representation. 

In the projector language, the quantum distances and 3-pt phases are written as
\bea 
\tr[P_V P_U] = \cos^2 \theta(u,v),\\
\arg \tr[P_V P_U P_W]  = \varphi(v,u,w).
\eea
The invariant tensors $g$ and $\omega$ arise the real and imaginary parts of a three-projector object \cite{avdoshkin2022}
\bea \label{eq:q-def}
\tr[P(x) \partial_{\alpha} P(x) \partial_{\beta} P(x)] = g_{\alpha\beta}(x) - \frac{i}{2} \omega_{\alpha\beta}(x),
\eea
known as the quantum geometric tensor (or Hermitian metric in mathematics). 
Here again $x$ denotes a coordinate on $\cpn$.

The generalization to higher-dimensional subspaces is straightforward. An $m$-dimensional subspace $\text{span}(|v_i\rangle)$ (an element of $Gr_{m,n}$, is  represented by
\bea
P_V = \sum_{i=1}^{m} | v_i \rangle \langle v_i |,
\eea
where we assumed that the $|v_i\rangle$ are orthonormal. Being an orthogonal projector, $P$ satisfies: $P_V^2 = P_V, ~ \text{tr} ~ P_V = m, P_V^{\dagger} = P_V$.

Finally, we note that under the action of $U(n)$ projectors  transform as
\bea
P_V \to U P_V U^{\dagger}.
\eea
This action is transitive and the stabilizer of a given $P_V$, i.e. the collection of $U$ such that $P = U P_V U^{\dagger}$, is $U(m) \times U(n-m) \subset U(n)$. Along with Eq. \eqref{eq:grasm_def} this completes the analogous dictionary in the Grassmannian case: points of $Gr_{n,m}$ are in one-to-one correspondence with the subspaces they represent.

\section{Subspace invariants} \label{sec:invariants}

Before finding the subspace invarinats explicitly, we would like to understand how many independent ones we expect. \if A set of $l$ $m$-dimensional subspaces can be though as an element of 

\bea
\underbrace{Gr_{m,n} \times \cdots \times Gr_{m,n} }_{\text{$l$ times}} = Gr_{m,n}^{\times l}.
\eea
We are now interested in the invariants of the diagonal action of $U(n)$ on this space. In other words, we wish to understand the structure of the space 

\bea
Gr_{m,n}^{\times l}/U(n).
\eea
\fi In Appendix \ref{app:dim}, we do this analysis and find the number of independent parameters to be 

\bea \label{eq:dim}
\begin{cases} m &,~\text{if} ~ l = 2\\ m^2 l (l - 2) + 1 &,~\text{otherwise} \end{cases},
\eea

where $l$ is the number of subspaces. The parameters are some real numbers which, for $n$ large enough, can vary independently of each other. We expect the parameters to obey some triangle inequalities and the independence might disappear for special configurations, prohibiting the configuration space from being a manifold. We will not discuss these subtleties here, nor will we discuss the topology of the configuration space. An interested reader can refer to the texts on invariant theory, e.g. \cite{mukai2003introduction}.

Instead, we will focus on interpreting the invariants enumerated in Eq. \eqref{eq:dim} from the linear algebra perspective.

\subsection{Two subspaces}

For a pair of one-dimensional subspaces  (lines) $V = {\rm span}\{| v \rangle\}$ and $W = {\rm span}\{| w \rangle\}$ there is a single invariant --- the angle \cite{avdoshkin2022}:
\bea \label{eq:angle}
|\langle v | w \rangle|^2 = \tr [P_V P_W] = \cos \theta^2(V, W),
\eea
meaning that given pair of vectors $|v\rangle, |w\rangle$ and $|v'\rangle, |w'\rangle$ with the same $\theta$, there is a $U(n+1)$ transformation that maps $|v\rangle$ to $|v'\rangle$ and $|w\rangle$ to $|w'\rangle$. 

For $m$-dimensional subspaces there are multiple invariants one can construct that generalize Eq. \ref{eq:angle} \cite{qiu2005unitarily}. They can be exhausted by the following iterative procedure. Let us find the vectors $v$ within $V_1 = V$ and $w$ within $W_1 = W$ that maximize the overlap:
\bea
\max_{v \in V_1, w \in W_1}|\langle v| w \rangle| = \cos \theta_1(V,W).
\eea
We will call the corresponding unique (up to a phase) vectors $v^W_1$ and $w^V_1$. Now we repeat this procedure in the orthogonal complements of $v_1$ and $w_1$: $V_2 = v_1^\perp $, $W_2 = w_1^{\perp}$. Repeating this procedure $m$ times we will end up with a pair of conjugate bases that, according to Theorem \ref{th:principal_vecs}, obey
\bea \label{eq:basis_constr}
\langle v^{W}_i | w^{V}_j \rangle = \delta_{i j} \cos \theta_i(V,W).
\eea
Note that we have also taken advantage of phase arbitrariness to make the overlap a non-negative real. Theorem \ref{th:pair} in the Appendix \ref{app:proofs} established $\theta_i$ as the complete set of invariants for a pair of subspaces. We refer to $\theta_i(V,W)$ as principal angles \cite{jordan1875essai, bjorck1973numerical} and the vectors $| v^{W}_i \rangle, |w^{V}_j \rangle$, satisfying Eq. \eqref{eq:basis_constr}
as principal vectors.
It is natural to ask whether the set of $\theta_i$ can be interpreted as different ways of defining distance on the Grassmannian. In Sec. \ref{sec:finsler}, we will show that (certain functions of) $\theta_i$'s are indeed distances computed with respect to Finsler metrics \cite{bao2012introduction}.

The same procedure can be performed in terms of projectors. Given two projectors $P_V$ and $P_W$, the principal angles can be obtained as
\bea \label{eq:proj_spec}
\{\cos^2 \theta_i(V,W)\} = \text{spec} P_V P_W P_V = \text{spec} P_W P_V P_W,
\eea

where $\text{spec}$ stands for the collection of eigenvalues, and one should think of $P_V P_W P_V$ ($P_W P_V P_W$) as an operator on $V$ ($W$). The projectors on the principal vectors $P_i^{V\to W} = | v^W_i \rangle \langle v^W_i |$ can also be obtained as polynomials of $P_V P_W P_V$.

\subsection{Three subspaces. Characterization via unitaries}
\label{sec:three_subspaces}

Let us consider three subspaces $V,W$ and $U$ . One can define three sets of principal angles associated with every pair of subspaces: $\theta_i(V,W)$, $\theta_i(W, U)$ and $\theta_i(V, U)$. In addition to that, we can define objects that involve all three subspaces.

\begin{figure}
\center
\includegraphics[width=0.5\textwidth]{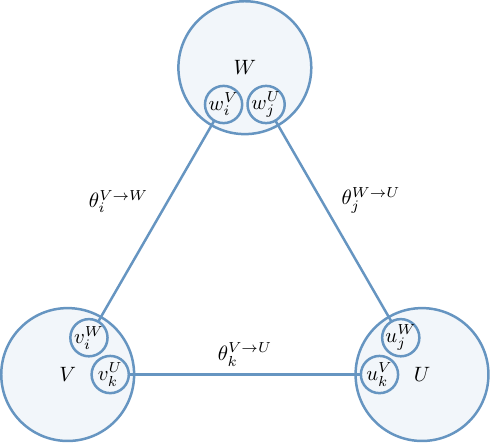}
  \caption{Graphic representation of 3-point invariants.}
   \label{fig:3-pt_invariants}
\end{figure}

Notice that now we can construct three pairs of conjugate bases Eq. \eqref{eq:basis_constr}, see Fig. \ref{fig:3-pt_invariants}. There are two distinct basis in each subspace and one can define a unitary matrix that connects them:

\bea \label{eq:unitaries}
\mathcal{V}^{U, W}_{i k} = \langle v^W_i | v^U_k \rangle, ~ \mathcal{W}^{V, U}_{i j} = \langle w^V_i | w^U_k \rangle, ~ \mathcal{U}^{V, W}_{j k} = \langle u^V_k | u^W_j \rangle.
\eea

There is an arbitrariness in defining these matrices that comes from the choice of phase in Eq. \ref{eq:basis_constr}, $ |w^V_i\rangle \to e^{i \phi_i }|w^V_i\rangle, |v^W_i\rangle \to e^{i \phi_i }|v^W_i\rangle $, ..., $3m$ such transformations altogether.

In App \ref{app:proofs}, we show that the unitaries Eq. \eqref{eq:unitaries} along with $\theta$'s determine the configuration uniquely. Furthermore, modulo phase ambiguity, parametrically, there is no further ambiguity. To see this, let us note that the number of independent gauge transformations is $\dim \left[U(1)^{3m}/U(1)\right]$ where the $U(1)$ in the denominator comes from the triviality of action when all phases are the same. Thus, adding up parameters in the matrices \eqref{eq:unitaries}, $m^2$ each, the three sets principal angles $\theta_i$ and subtracting the gauge transformations, we obtain
\bea
3 m^2 + 3 m - (3m-1) = 3 m^2 + 1,
\eea
which agrees with Eq. \eqref{eq:dim} for $l=3$.

Theorem \ref{th:gps_2} in Appendix \ref{app:proofs} additionally shows if we have a collection of $l$ subspaces their configuration is uniquely determined by the principal angles computed for any pair of subspaces and the unitaries in Eq. \eqref{eq:unitaries} computed for every triple. In this case, we can also identify the independent variables. Now, we have $l-1$ induced bases in each subspace. One can define a principal unitary matrix for every pair of bases, but only $l-2$ of them are independent (multiplying them circularly yields identity). Thus, we have $l(l-2)$ independent unitary matrices, each characterized by $m^2$ parameters. Additionally, we have $m\cdot l(l-1)/2$ distances that we need to add to the number of invariants and $m\cdot l(l-1)/2 - 1$ gauge-transformations that need to be subtracted, leading to Eq. \eqref{eq:dim}.

We have established that the principal angles along with the principal unitaries give a complete description of a configuration of subspaces. However, the unitaries are not invariant (under phase changes) and, thus, cannot be reduced to projectors on the corresponding subspaces. 
In the next section, we find a description that is invariant.

\subsection{Three subspaces. Invariant characterization}

Here we will define three classes of invariant objects that contain the same information as the unitaries Eq. \eqref{eq:unitaries}. However, unlike the unitaries, these objects are not fully independent from each other and should satisfy certain relations. Deriving these relations is beyond the scope of this work.

We begin with 2-state overlaps (2SO), given by the absolute values of the overlaps matrices:
\bea \label{eq:overlaps}
\cos^2 \theta_{ik}(V;W,U) = |\langle v^{W}_i | v^{U}_k \rangle|^2,~ \cos^2 \theta_{ij}(W;V,U) = |\langle w^{V}_i | w^{U}_j \rangle|^2,~ \cos^2 \theta_{jk}(U;V,W) = |\langle u^{V}_j | u^{W}_k \rangle|^2.
\eea
These quantities can be computed in the projector formalism as well. For that one needs to look at the spectra of $P_V P_W P_V$ and $P_V P_U P_V$ and take the overlaps of $i$-th and $k$-th eigenvectors. This information can be extracted from objects of the form $\tr[(P_V P_W P_V)^p (P_V P_U P_V)^q]$, for all natural $p$ and $q$.

The second class is the 4-state phases (4SP):
\bea \label{eq:4pt_phases}
\varphi_{j j' k k'}(V;W,U) = \arg \langle v^W_i | v^U_k \rangle \langle v^U_k | v^W_{i'} \rangle \langle v^W_{i'} | v^U_{k'} \rangle \langle v^U_{k'} | v^W_i \rangle.
\eea
 The 4-pt phases reduce to
\bea
\tr[(P_V P_W P_V)^l (P_V P_U P_V)^k (P_V P_W P_V)^{l'} (P_V P_U P_V)^{k'}].
\eea
Both 2SO and 4SP come from a single principal unitary and became trivial for one-dimensional subspaces. The last class of invariants requires all three unitaries to be constructed. These invariants generalize the Pancharatnam phase and we will refer to them as 3-state phases (3SP). 3SP are given by
\bea \label{eq:3-pt-phase}
\varphi_{ijk}(V,W,U) = \arg \langle v^{U}_k | v^{W}_i \rangle \langle w^{V}_i | w^{U}_j \rangle \langle u^{W}_j | u^{V}_k \rangle,
\eea
or, in the projector representation,
\bea 
\varphi_{i,j,k}(V,W,U) = \arg \tr [P^{V\to W}_i P^{W\to V}_i P^{W \to U}_j P^{U \to W}_j P^{U\to V}_k P^{V\to U}_k],
\eea
where the projectors $P^{V \to W}_i$ were introduced after Eq. \eqref{eq:proj_spec}. In the one-dimensional case, the only possible value for indices is $i=j=k=1$ and Eq. \ref{eq:3-pt-phase} reduces to the Pancharatnam phase.

In Appendix \ref{app:unitaries}, we prove that the 4SP and 3SP contain all gauge-information in the phases of principal unitaries. Along with 2SO this gives a complete invariant characterization of a configuration of subspaces and, thus, we have answered Question 1.

\section{Local geometric structures} \label{sec:tensors}

In this section, we introduce some local geometric structures on $Gr_{m,n}$ and show how they can be used to compute the subspace invariants introduced earlier.

As a natural generalization of the quantum geometric tensor Eq. \ref{eq:q-def} we define its matrix valued version \cite{ma2010abelian}

\bea
Q_{\alpha \beta}(x) = P(x)\partial_{\alpha}P(x)\partial_{\beta}P(x)P(x),
\eea

where for fixed $\alpha$ and $\beta$, $Q_{\alpha \beta}$ should be thought as a matrix on the subspace represented by $P(x)$, i.e. an $m\times m$ matrix.
Mathematically, $Q$ is a rank-2 tensor on the tangent space of $Gr_{m,n}$ that takes values in fiber endomorphisms (here fiber refers to the tautological bundle on $Gr_{m,n}$).

We can decompose it into the symmetric and anti-symmetric (in terms of the tangent indices) or, equivalently, Hermitian and anti-Hermitian (in terms of the fiber) parts:
\bea \label{eq:g_and_f}
Q_{\alpha \beta}(x) = G_{\alpha \beta}(x) - \frac{i}{2} F_{\alpha \beta}(x).
\eea
We will refer to the symmetric part, $G_{\alpha \beta}(x) = \frac{1}{2}(Q_{\alpha \beta}(x) + Q_{\beta \alpha}(x)) = -\frac{1}{2}P(x) \partial_{\mu} \partial_{\nu} P(x) P(x)$, as the matrix-valued metric tensor or the $G$-tensor. Formally, it is a structure on the tautological bundle of $Gr_{m,n}$, but we could not find any mathematics literature that discusses it. In the following sections, we show how $G$ defines several notions of distance and some 3-point objects on the Grassmannian manifold. We will not develop a general theory of $G$-tensors in this work.

The anti-symmetric part 
\bea \label{eq:f-def}
F_{\alpha \beta}(x) = \frac{1}{2}(Q_{\alpha \beta}(x) - Q_{\beta \alpha}(x)) = \partial_{\alpha} A_{\beta} - \partial_{\beta} A_{\alpha} + A_{\alpha}\cdot A_{\beta} - A_{\beta}\cdot A_{\alpha},
\eea
is the curvature of the Wilczek-Zee (tautological) connection $(A_{\alpha})_{ij} =  i \langle \psi_i| \partial_{\alpha} \psi_j \rangle$ introduced in \cite{PhysRevLett.52.2111}. Connections on vector bundles  and the corresponding curvatures are well known in mathematics \cite{dubrovin2012modern}. The role of $A$ and $F$ in physics has been recognized since the work of Wilzek and Zee \cite{PhysRevLett.52.2111} and the physics community is familiar with the mathematical formalism of non-Abelian connections due to the significant role they play in the Standard model of particle physics \cite{peskin1995introduction}.

There is an important subtlety that it is the connection $A$ that arises in calculations with degenerate states and, in the non-abelian case, $A$ cannot be reconstructed from its curvature, even locally \cite{PhysRevD.12.3843}. Because of that, $A$ is a more fundamental object and $F_{\mu\nu}$ alone is not sufficient to capture all the geometry coming from the $A$. This is contrary to the Abelian case, where, as we saw in Sec. \ref{sec:one_d}, $\omega$ could be used to determine the 3-state phases due to Stokes' theorem. Thus, in what follows we will use $A$ and $G$ as the two basic objects that all geometry can be reduced to. 

We expect \cite{DESER1979189} that the extra information in $A$ that is not captured in $F$ can be recovered from the covariant derivatives of $F$, e.g. $\nabla^{A}_{\alpha} F_{\beta\gamma}= \partial_{\alpha} F_{\beta\gamma} - [A_{\alpha}, F_{\beta\gamma}]$, $\nabla^{A}_{\alpha} \nabla^{A}_{\alpha'} F_{\beta\gamma}$,..., where here $[,]$ stands for commutator. Under a basis change $U$ in the fiber, the connection transforms as $A_{\alpha} \to U^{\dagger} A_{\alpha} U + U^{\dagger} \partial_{\alpha} U$, whereas $G, F$ and the derivatives of $F$ ($\nabla^{A}_{\alpha} F_{\beta\gamma}, \dots$) are co-variant tensors, they transform as
\bea
G(x), F(x) \to U^{\dagger}(x) G(x) U(x), U^{\dagger}(x) F(x) U(x)
\eea
under a basis change $U$ in the fiber. Strictly speaking, only their traces are invariant. However, working with co-variant matrix-valued objects is just a choice of convenience and one could instead define a set of invariant scalars of the form $\tr[G_{\alpha\beta} G_{\mu \nu}]$, $\tr[F_{\mu\nu} G_{\lambda \rho}]$, $\tr[F_{\mu\nu} \cdots F_{\lambda \rho} \nabla^{A}_{\alpha} F_{\beta\gamma}], \dots$ (traces of all combinations of $F, \nabla^{A} F, \dots$ and $G$) that have more tangent indices. One usually chooses to work with co-variant objects because it simplifies algebra and it is the conventional approach to non-Abelian objects, at least in the theoretical physics literature \cite{peskin1995introduction}. The same applies to working with $A$ instead of $F$ and its derivatives. When deriving the invariants associated to a collection of points on $Gr_{m,n}$ we will always obtain true invariants. Roughly speaking, it corresponds to taking traces of combinations of $G$ and $F$ (and its derivatives) in all final expressions. 

In the rest of this section we will reduce all other geometric objects to $A$ and $G$ thus establishing them as the local geometric structures that we asked for in Question 2.

\subsection{The matrix-valued metric tensor $G$ and Finsler metrics}\label{sec:finsler}

We refer to $G$ as the matrix-valued metric tensor, because it can be use to define several metric tensors. First of all, we can define the standard Riemannian metric as
\bea\label{eq:riehmann_g}
g^{(1)}_{\alpha \beta}(x) = \tr[G_{\alpha \beta}(x)].
\eea
It can be called the Fubini-Study metric because it arises as the pull-back of the Fubini-Study metric on $\cpn$ under the Pl\"ucker embedding, see App \ref{app:pluecker}. This also allows one to see the geometric structure represented by $G$ as a generalization of Riemannian structure. In the case $m>1$, $G$ contains additional information and one can form fully-symmetric tensors of higher ranks:

\bea \label{eq:finsler_def}
g^{(2)}_{\alpha_1 \beta_1 \alpha_2\beta_2} = \tr[G_{\{\alpha_1 \beta_1} G_{\alpha_2 \beta_2\}}],~\dots~, g^{(l)}_{\alpha_1 \cdots \beta_{l}} = \tr[G_{\{\alpha_{1} \beta_{1}} \cdots G_{\alpha_{l} \beta_{l}\}} ] 
,~ \dots
\eea
where the square brackets in indices stand for symmetrization. Fully-symmetric tensors like $g^{(l)}$ are known as Finsler metrics \cite{bao2012introduction} and can be used to define length according to

\bea
\mathcal{L}^{(l)}(\gamma)=\int_{\gamma} \sqrt[2l]{g^{(l)}_{\alpha_1 \cdots \beta_{l}} dx^{\alpha_1} \cdots dx^{\beta_{l}}},
\eea
where $\gamma$ is the curve for which we are computing the length. In \cite{HU201118}, it was shown that the Finsler metrics on $Gr_{m,n}$ with $l = 1,\text{min}(k,n-k)$ are functionally independent.

It is natural to expect that the independent Finsler metrics on $Gr_{k,n}$ lead to principal angles. Given two points on $Gr_{m,n}$ that correspond to subspaces $V$ and $W$, we might ask what is the length of the shortest path that connects them. This path is known as a geodesic and in App. \ref{app:geodesics} we find that all Finsler metric $g^{l}$ share the same geodesic, that we will denote as $\gamma(V,W)$. At the same, different $g^{(l)}$ assign different lengths to $\gamma(V,W)$, explicitly

\bea
\mathcal{L}_l(V,W) = \sqrt[2l]{\sum^{m}_{i=1} \theta_i^{2l}(V,W)},
\eea
where $\theta_i(V,W)$ are the principal angles and the details of the calculation are delegated to App. \ref{app:geodesics}. Knowing $\mathcal{L}_l(V,W)$ for $l=1,\cdots,m$ is equivalent to knowing the principal angles, and we conclude that the tensor $G$, or more specifically the Finsler metrics $g^{(l)}$, can be used to compute them.

The Finsler metrics do not capture the full geometric information contained in $G$. Roughly, this is the information that is lost when we symmetrize the traces in Eq. \eqref{eq:finsler_def}. In the next section, we show that $G$ can also be used to determine some of the 3-subspace invariants.

\subsection{Two-state overlaps and 4-state phases}

In the previous section, we saw how the $G$ tensor defines a unique geodesic given two points and assigns $m$ different lengths to it. Now we show how one can extract some information about how the subspace is changing along the geodesic. This will allow us to compute the 2SO from Eq. \eqref{eq:overlaps} and 4SP from Eq. \eqref{eq:4pt_phases}. 

For two subspaces $V$ and $W$, we denote as $t_{\alpha}^{V \to W}$ the tangent vector to the geodesic at $V$. Next, using the results from App. \ref{app:geodesics}, we find
\bea
G(V\to W) \equiv t^{V\to W}_{\alpha} t^{V\to W}_{\beta} G_{\alpha\beta} = \sum_i (\theta^{VW}_i)^2 | v^W_i \rangle \langle v^{W}_i |,
\eea
where one can also use the right-hand side to define the normalization of $t_{\alpha}^{V \to W}$.
Combining the first $m$ powers of $G(V\to W)$, generically, we can isolate the projectors on all principal vectors $P_i^{V\to W}=| v^W_i \rangle \langle v^{W}_i |$.

Given three subspaces $V$, $U$ and $W$, we can construct invariants out of both $G(V\to W)$ and $G(V \to U)$ or, equivalently, out of $| v^W_i \rangle \langle v^{W}_i |$ and $| v^U_k \rangle \langle v^{U}_k |$. This means that the 2SO Eq. \eqref{eq:overlaps} are reconstructible from traces of the form
\bea
\tr[G^p(V\to W) G^q(V\to U)],
\eea
where $p,q = 1,\cdots,m$. Similarly, the other 2SO are extracted from $\tr[G(W\to V)^l G^f(W\to U)]$ and $\tr[G(U\to V)^l G^f(U\to W)]$.

Analogously, 4SP Eq. \eqref{eq:4pt_phases} can be computed from objects of the form $\tr[G(V\to W)^{l_1} G^f(V\to U)^{f_1}G(V\to W)^{l_2}G^f(V\to U)^{l_2}]$, where $l_1, l_2 ,f_1, f_2$ are again non-negative integers. 

Due to the fact that 2SO and 4SP involve only a single principal unitary we only took products of vectors within the same subspace and never needed to use the connection $A$.

\subsection{3-state phases}

The last invariants that we need to compute are the 3SP from Eq. \eqref{eq:3-pt-phase} and for them we will have to use both the $G$-tensor and $A$. The main ingredient in this construction are the U(1) bundles one can define over  geodesics corresponding to the change in the $i$-th principal vector.

Consider the geodesic $\gamma(V,W)$ and let us track the $i$-th principal vector defined by Eqs. \eqref{eq:v_change} and \eqref{eq:phi_change} in App. \ref{app:geodesics}. Since, we have a one dimensional space over every point of the geodesic, this defines a $U(1)$-bundle and its connection is given by $A^i_{\alpha}(V\to W)(t) = -i\langle v_i(t) |\partial_{\alpha}v^i(t)\rangle$, where $t$ is the a coordinate along the geodesic. This U(1)-bundle is a subbundle of the tautological $U(m)$ bundle which is obtained by restricting it on the $P_i(t)$ subspace. The integral of the connection $\int_{\gamma(V,W)} A^i(V\to W)$ (we are using the differential form notation here) gives us the phase change of $P_i(t)$ along $\gamma(V,W)$. Since $\gamma(V,W)$ is not a loop, the integral is not gauge-invariant and depends on the choice of representative vectors at the ends of $\gamma(V,W)$. This arbitrariness will go away when we combine it with other gauge-non-invariant objects in the final expression.

Let us include the third subspace $U$, and consider the geodesic $\gamma(V\to U)$. We can again compute $G(V\to W)$ and $G(V\to U)$ and, diagonalizing  them, get access to pseudo-principal vectors $|\tilde{v}^W_i \rangle$ and $|\tilde{v}^U_k \rangle$, which differ from the true principal vectors satisfying Eq. \eqref{eq:basis_constr} by arbitrary phases. We can, however, choose the phase of $| \tilde{v}^W_i \rangle$ to be consistent with the one used in calculating the integral in the previous paragraph. This makes the expression $\langle \tilde{v}^U_k|\tilde{v}^W_i \rangle \exp\left(\int_{\gamma} A^i(V\to W)\right)$ insensitive to the phase of $| \tilde{v}^W_i \rangle$. Combining the geodesics and vector overlaps in the other parts of the geodesic triangle, we obtain a expression for the 3SP Eq. \eqref{eq:3-pt-phase}:

\bea
\varphi_{i,j,k}(V,W,U) = \langle \tilde{v}^U_k|\tilde{v}^W_i \rangle \exp\left(i\int_{\gamma} A^i(V\to W)\right) \langle \tilde{w}^V_i | \tilde{w}^U_j \rangle \exp\left(i\int_{\gamma} A^j(W\to U)\right)\langle \tilde{u}^W_j | \tilde{u}^V_k \rangle \exp\left(i\int_{\gamma} A^k(U\to W)\right),
\eea

where $j$ and $k$ stand for the choices of other principal vectors. We stress again, that the difference from Eq. \eqref{eq:3-pt-phase} comes from working with pseudo-principal vectors rather than the true principal vectors that satisfy Eq. \eqref{eq:basis_constr}. Effectively, we use the connection $A$ to compensate for that.

The construction in this section only uses the local structures $G$ and $A$ that live in the tautological bundle. In particular, we only take overlaps of vectors from the same fiber and comparing vectors from different fibers always involves the connection $A$. This should be contrasted with the considerations in Sec. \ref{sec:three_subspaces}, where we could compute overlaps of vectors from different subspaces (fibers). The results of this section answer Question 3 and conclude our construction.

\section{Outlook}
In this work, we have described the invariants associated with a collection of subspaces of a finite-dimensional Hilbert space and given their interpretation in terms of local geometric structures on the Grassmannian. These invariants generalize the concepts of quantum distance and Pancharantman phase to $m$-dimensional subspaces. 

An unexpected outcome of this construction is the significant role played by the $G$-tensor, defined in Eq. \eqref{eq:g_and_f}, in calculating the invariants. For example, despite its "metric-like" nature, it can be used to determine the 4SP, Eq. \eqref{eq:4pt_phases}. We believe the $G$-tensor deserves to be studied more thoroughly. In that regard, we would like to highlight two naturally-emerging mathematical questions.

\textbf{Question 1:}
{\textit{What are the general properties of "$G$-manifolds", i.e. fiber bundles with a $G$-tensor?} They provide a special example of Finsler manifolds, with $m$ Finsler metrics that originate from the same structure. One can wonder if all $m$ Finsler metrics would still have the same geodesics as we saw for the Grassmannian. $G$-manifold structure arises in physics when one has a degenerate state or, more generally a subspace, that depends on a parameter. In this case, the $G$-tensor is the pullback on the parameter space of the $G$-tensor on $Gr_{m,n}$ that we described in the present work. In this regard, we can also ask if all $G$-manifolds arise this way, similarly to how it works for vector connections \cite{universal-connection}.}
\newline

\textbf{Question 2:}
{\textit{How does the $G$-tensor relate to the connection $A$ and the complex structure on $Gr$?} On $\cpn$, $g$ and $\omega$ can be thought of as coming from the K\"ahler structure. One can wonder if it is possible to define a non-abelian version of the K\"ahler structure that combines $G$ and $A$ with the complex structure on $Gr_{m,n}$.}
\newline

The physical significance of the mathematical structures described in this work can be most directly illustrated by a particular type of non-adiabatic evolution. It is a version of the bang-bang protocol \cite{PhysRevA.58.2733}, where the changes in the Hamiltonian are followed by projection on a subspace. Let the system be originally prepared in the state $|\psi\rangle$ from a degenerate subspace of some Hamiltonian $H(\lambda)$ and we change $\lambda$ several times such that the degenerate subspace changes without destroying the degeneracy. If we additionally eliminate all excited states, the resulting state will be given by $P_{n} \cdots P_2 P_1 |\psi\rangle$ (up to an overall dynamical phase). Using the approach developed in this work, one can derive a concise description of this evolution in terms of geometric objects. 

Besides that, our formalism can help understand various other systems with degeneracy in the spectrum. Band structures of materials with both PT-symmetry and spin-orbit coupling exhibit exact two-fold degeneracy at all values of the quasi-momentum due to Kramers degeneracy. We expect the much richer geometry described here to replace the standard Berry curvature and quantum metric when computing the properties of such materials. Understanding anyon braiding beyond topology, e.g. finding optimal trajectories, susceptibility to errors, can be aided by this formalism. Finally, for systems where some quantities are determined by the non-Abelian Berry curvature, we expect some other quantities to be determined by the $G$-tensor (or both $G$ and $F$). For instance, in axion insulators, the magnetoelectric polarizability is given by the integral of the Chern-Simons form and it would be interesting to find a related response that is also sensitive to the $G$-tensor.

\section{Acknowledgments}

We thank Joel Moore, Nikita Sopenko, Yakov Kononov, Alexey Milekhin, \"Omer Aksoy, Ivan Karpov and especially Fedor Popov and Daniel Chupin for many helpful discussions. We also thank Johannes Mitscherling, Daniel Chupin, Max Geier, Oriana Diessel and Andrey Grekov for help with the manuscript and Johannes Mitscherling, Dan Borgnia, Max Geier, Yugo Onishi and Liang Fu for collaboration on related projects. 
The author was supported by a Kavli ENSI fellowship and National Science Foundation through QLCI grant OMA-2016245 during his time at UC Berkeley and the National Science Foundation (NSF) Convergence Accelerator Award No. 2235945 at MIT.

\newpage

\appendix

\section{Parameter counting} \label{app:dim}

If we have a manifold $M$ that is acted upon by a group $G$ (not to be confused with the matrix-valued metric tensor $G$ used in main text), we define the number of “invariants” as the co-dimension of the orbit of a generic point $p$ of $M$ under $G$. The dimension of the orbit is given by $\dim G - \dim Stab_p$, where $Stab_p \subset G$ is the stabilizer subgroup of $p$.

Let us start with the 1D case. For $l$ subspaces, the manifold in question is
\bea
M = \underbrace{\cpn \times \cdots \times \cpn}_{\text{$l$ times}},
\eea
the group $G = U(n+1)$ and its action is diagonal. The stabilizer of a generic point (where all $l$ lines are linearly independent) is 
\bea
Stab = U(1) \times U(n + 1 - l),
\eea
leading to 
\bea \label{eq:cpn_count}
\dim_{\mathbb{R}} M/U(n+1) = 2 l n - (n+1)^2 + 1 + (n+1-l)^2 = (l-1)^2.
\eea
In particular, for $l=2$ we have $1$ invariant corresponding to the quantum distance, for $l=3$ we have 4 corresponding to the 3 pair-wise quantum distances and one 3-pt phase. For arbitrary $l$, the invariants are the $\frac{l(l-1)}{2}$ pair-wise distances and $\frac{(l-1)(l-2)}{2}$ 3-pt phases (one only needs to consider the ones where one of the points is kept the same) which agrees with Eq. \eqref{eq:cpn_count}.

The calculation for the Grassmannian is analogous. For $l$ subspaces the manifold is
\bea
M = \underbrace{Gr_{m,n} \times \cdots \times Gr_{m,n}}_{\text{$l$ times}}
\eea
and $G = U(n)$. In what follows, we assume that the dimension $n$ of the ambient space $\mathbb{C}^n$ is large enough for the subspaces to be linearly independent, i.e. $n > m l$. 
The stabilizer has slightly different forms for $l=2$ and all other values of $l$. For $l=2$, we have
\bea
Stab = U(1)^{m} \times U(n - 2m).
\eea
For number of invariants this yields
\bea
\text{dim}~ M/U(n) = 4 m (n-m) - n^2 + m + (n-2m)^2 = m
\eea
and this corresponds to the principal angles that we construct in Sec. \ref{sec:invariants}.

For all other $l$,
\bea \label{eq:l_subspaces}
Stab = U(1) \times U(n - l \cdot m)\nonumber\\
\text{dim}~ M/U(n) = 2 l m (n-m) - n^2 + 1 + (n-l m)^2 = m^2 l (l - 2) + 1.
\eea
In particular, for $l=3$ we have $3m^2 + 1$, and we explain the meaning of the new invariants in Sec. \ref{sec:three_subspaces}. 

\section{Pl\"ucker embedding} \label{app:pluecker}

The Pl\"ucker embedding is a representation of an $m$-dimensional subspace $\mathbb{C}^n$ as a one-dimensional subspace of the exterior power $\Lambda^m $ of $\mathbb{C}^n$. It corresponds to constructing the Slater determinant in physics. Below we overview the structure of the Pl\"ucker embedding and its relation to the geometric invariants described in this work.

The Pl\"ucker embedding is a map $Pl: Gr_{m,n} \to P(\Lambda^m \mathbb{C}^n)$, where $P$ stands for projectivization. It is defined by

\bea
Pl: V = \text{span}(v_1, \dots, v_m) \to \text{span}( v_1 \wedge \dots \wedge v_m),
\eea
where $\wedge$ stands for exterior product. For two subspaces, $V = {\rm span} \{v_i\}$ and $W = {\rm span} \{w_i\}$, the inner product that is induced on the exterior power is simply

\bea
\langle Pl~ V, Pl~ W \rangle:= \det \left\{ \langle v_i| w_j\rangle \right\}.
\eea

Given three $m$-dimensional subspaces  $V = {\rm span} \{v_i\}$, $W = {\rm span} \{w_i\}$ and $U = {\rm span} \{u_i\}$ we define the three-point function:

\bea
P(V, W, U)=\langle Pl~ V, Pl~ W \rangle 
\langle Pl~ W, Pl~ U \rangle 
\langle Pl~ U, Pl~ V \rangle = \det \left\{\langle v_i | w_j \rangle\right\} \det \left\{\langle w_j | u_k \rangle \right\} \det \left\{\langle u_k | v_i \rangle \right\},
\eea
where we sum over $l$ and $m$ indices and the $\det$ is taken over $i$ and $j$ indices. Alternatively, we can write this invariant as

\bea
P(V, W, U) = \det \left[ P_V P_W P_U P_V \right],
\eea

where $\det$ is taken over the $V$ subspace. By Taylor expanding $P(V, W, U)$ we arrive at the quantum geometric tensor

\bea
Q^{\text{Pl\"ucker}}_{\alpha \beta}(x) = \tr [Q_{\alpha \beta}(x)].
\eea

The tensor $Q^{\text{Pl\"ucker}}$ represents the abelian part of the geometry of the $Gr$: its real part is the Fubini-Study metric $g^{(1)}$ from Eq. \eqref{eq:riehmann_g} and its imaginary part is the $U(1)$ part of the non-aberlian connection Eq. \eqref{eq:f-def}.

\section{Distances and geodesics}
\label{app:geodesics}

Let us find the geodesic between two points on the Grassmannian corresponding to subspaces $V$ and $W$. First of all, there are $m$ distinct way of computing distance, namely by integrating

\bea
\sqrt[2l]{\tr\left\{\left[ P(x)  \left(\frac{dx^{\mu}}{d\tau} \partial_{\mu}P(x)\right)^2\right]^l\right\}}
\eea
along a line with $l$ running from $1$ to $\min(m,n-m)$.

Any line that connects $V$ and $W$ without leaving ${\rm span}(V,W)$ can be represented as

\bea
P(t) = \sum_i |v_i(t)\rangle \langle v_i(t)|.
\eea

\bea \label{eq:v_change}
|v_i(t)\rangle = \cos(\varphi(t))| v_i \rangle + \sin(\varphi(t)) | w_i^{\perp}\rangle,
\eea
where we have defined $| w_i^{\perp}\rangle = \frac{| w_i\rangle - \cos \theta_i | v_i\rangle}{\sin \theta_i}$ and the boundary conditions are $\varphi_i(0) = 0$, $\varphi_i(1) = \theta_i$.

We have $|\dot{v}_i(t)\rangle = \dot{\varphi}_i( -\sin(\varphi(t))| v_i \rangle + \cos(\varphi(t)) | w_i^{\perp}\rangle $) and 

\bea
\dot{P} = \sum_i | \dot{v}_i \rangle \langle v_i | + | v_i \rangle \langle \dot{v}_i |
\eea

This leads to

\bea
\left(\tr\left\{\left[ P(x)  \left(\dot{P}(x)\right)^2\right]^l\right\}\right)^{1/2l} = \left(\sum_i \langle v_i | v_i \rangle^l\right)^{1/2l} = \left(\sum_i (\dot{\varphi}^2)^l \right)^{1/2l}
\eea

Thus for the length functional we have

\bea
\mathcal{L}_l = \int_0^1 dt \left(\sum_i \dot{\varphi}^{2l} \right)^{1/2l}
\eea

varying it leads to

\bea
\delta \mathcal{L}_l = \int_0^1 dt \frac{ \sum_i \dot{\varphi}_i \dot{\varphi}_i^{2l-1} }{\left(\sum_j \dot{\varphi}_j^{2l} \right)^{1/2l}} = -\int_0^1 dt \sum_i \varphi_i \frac{d}{dt} \left(\frac{ \dot{\varphi}_i^{2l-1} }{\left(\sum_j \dot{\varphi}_j^{2l} \right)^{1/2l}}\right).
\eea

Going to a parametrization of the curve with $\sum_j \dot{\varphi}_j^{2l} = const$, we conclude that $\ddot{\varphi}_i = 0$ for all $l$. Along with the boundary conditions, this leads to the unique solution:

\bea \label{eq:phi_change}
\phi_i = t \theta_i.
\eea

One should remark that the shape of the geodesic is independent of which Finsler metric one uses. Now we compute the lengths of that geodesic with respect to the different Finsler metric

\bea
\mathcal{L}_l = \sqrt[2l]{\sum_i \theta_i^{2l}}.
\eea

The knowledge of $\mathcal{L}_m$ is equivalent to the knowledge of $\theta_i$'s for all $i$.

\section{Invariant information in the principal unitaries} \label{app:unitaries}

Since the absolute values of the principal unitaries are invariant, we only focus on the phases of the matrix elements: $\varphi^{1}_{ik} = \arg \langle v^W_i | v^U_k \rangle$, $\varphi^{2}_{ij} = \arg \langle w^V_i | w^U_j \rangle$ and  $\varphi^{3}_{ki} = \arg \langle u^W_i | u^V_k \rangle$. They satisfy $\varphi^{q}_{ij} = - \varphi^{q}_{ji}$, for $q=1,2,3$. With this notation, the phase arbitrariness is captured by the following gauge transformation:
\bea \label{eq:conn_def}
\varphi^{1}_{ik} \to \varphi^{1}_{ik} + \alpha_i - \gamma_k,\\
\varphi^{2}_{ji} \to \varphi^{2}_{ji} + \beta_j - \alpha_i,\\
\varphi^{3}_{kj} \to \varphi^{3}_{kj} + \gamma_k - \beta_j.
\eea
In the physics language $\varphi^{q}_{ij}$ define a $U(1)$ gauge field on a complete 3-partite graph. We are going to prove that all gauge-invariants are generated by the circulation of certain 3- and 4- loops. We make these notions precise in the following

\textbf{Definitions.} Consider a directed graph $G$ with a vertex set $V$ and a directed edge set $E$. A $U(1)$ \textit{connection} on $G$ is a map from E to $U(1)$, i.e. we assign a phase to every edge of the graph. We can represent a connection as $\varphi_{ij}$ where $i,j$ corresponds to a pair of vertices connected by an element of $E$. We say that two $U(1)$ connections, $\varphi_{ij}$ and $\varphi'_{ij}$, are \textit{equivalent} if there exists a map $V \to U(1)$, that we denote $\alpha_i$, such that $\varphi'_{ij} = \varphi_{ij} + \alpha_i - \alpha_j$. Finally, for a loop $\sigma$ in graph $G$ with connection $\varphi$, the \textit{circulation} (\textit{Wilson loop}) $W(\sigma)$ of $\sigma$ is defined as $W(\sigma) = \sum\limits_{i,j\in \sigma} \varphi_{ij}$.

With these definitions we can prove the following

\begin{theorem}
Every equivalence class of $U(1)$ connections is uniquely determined by the circulation of the connection on loops in the graph (Wilson loops).
\end{theorem}
\textit{Proof:} We show that the circulations determine the connection uniquely in a particular gauge. The gauge is defined by choosing the connection to be zero on the edges of some spanning tree. Then adding any other edge in the graph to the spanning tree creates a loop. In this gauge, only one edge in this loop has a non-vanishing connection, so the value of the circulation is given by the connection on the corresponding edge. \textit{Q.E.D.}

Now, we need to understand the structure of loops in the 3-partite graph defined by Eq. \ref{eq:conn_def}. First, we introduce a

\textbf{Definition.} For two loops $\sigma$ and $\sigma'$, their \textit{sum}, denoted $\sigma + \sigma'$, is defined as the union of directed edges in both $\sigma$ and $\sigma'$, with the removal of edges and their opposites that appear together.

As a warm-up we will describe the loops on complete 2-partite graphs.

\begin{theorem}
In a complete 2-partite graph on $2n$ vertices, the loop space is generated by 4-loops of the form $\{e_{ij},e_{ji'},e_{i'j'},e_{j'i}\}$, where $i,i'$ and $j,j'$ index the vertices in the first and second parts, respectively. \end{theorem}

\textit{Proof:} Consider an arbitrary loop of length $2n$, $L_{2n}$, explicitly it can be represented as $i_1 j_1 i_2 j_2 \cdots i_n j_n$. The $L_{2n}$ is given by the sum of a length $2n-2$ loop $i_1 j_2 \cdots i_n j_n$ and a 4-loop $i_1 j_1 i_2 j_2$. Repeating this procedure $n-2$ times will give us a reduction of $L_{2n}$ to 4-loops of the desired form. \textit{Q.E.D.}

\begin{theorem}
In a complete 3-partite graph on $n+n+n$ vertices, the loop space is generated by 4-loops of the form $\{e_{ij},e_{ji'},e_{i'j'},e_{j'i}\}$ and 3-loops of the form $\{e_{ij},e_{jk},e_{ki}\}$ . \end{theorem}
\textit{Proof:} Let us consider an arbitrary loop $L$ of length $n$. Now we show how to reduce it to the fundamental loops of the desired form. We loop at the first three element of the loop, if they belong to only two parts, we do not change it. If they belong to different parts, i.e. the loop has a form $i_1 j_1 k_1 \cdots$, we can represent it as a sum $i_1 k_1 \cdots$ + $i_1 j_1 k_1$. This gives us a loop of length $n-1$. Next look at the following three consecutive vertices and repeat the procedure. At the end we have two options. Either we obtain a loop where every three consecutive vertices belong to only two parts, thus, giving us a loop in a two-partite graph, which according to theorem 2 reduces to 4-loops. Or we end up with a single 3-loop, in either case this gives us a required representation of the loop. \textit{Q.E.D.}

\section{Main theorems}\label{app:proofs}

We begin with proving a theorem that defines principal angles.

\begin{theorem} \label{th:principal_vecs}
Let $V, W$ be two $m$-dimensional subspaces of an $n$-dimensional vector space $\mathbb{C}^n$, then one can always find orthonormal bases $|v_i\rangle$ in $V$ and  $|w_j\rangle$ in $W$ such that, 
\bea\label{eq:basis_in_proof}
\langle v_i|w_j\rangle = \cos\theta_i ~\delta_{ij},
\eea
with $\cos\theta_i\geq 0$.
\end{theorem}

\textit{Proof:} Let us pick any basis $|v'_{i'}\rangle$ in $V$ and any basis $|w'_{j'}\rangle$ in $W$. Next, we perform the singular value decomposition of  $\langle v'_i| w'_j\rangle$:

\bea
\langle v'_i| w'_j\rangle = V_{ii'} D_{ij} W_{j'j},
\eea
where $V_{ii'}$ and $W_{j'j}$ are unitary matrices, $D_{ij}$ is diagonal and sums over repeated indices are implied. Then $|v_i \rangle = V_{ii'} |v'_{i'}\rangle$, $|w_j \rangle = W^{\dagger}_{j'j} |w'_{j'}\rangle$ and $\cos \theta_i = D_{ii}$ satisfy Eq. \eqref{eq:basis_in_proof} \footnote{We thank Mike Zalatel for pointing this proof out to us.}. \textit{Q.E.D.}

Next, we copy a statement from appendix B of \cite{avdoshkin2022},

\begin{theorem} \label{th:old}
For two sets of normalized vectors $|u_i\rangle$ and $|v_i\rangle$ such that $\langle u_i| u_j\rangle = \langle v_i| v_j\rangle$  there exists a unitary matrix $U$ that maps $|u_i\rangle$ and $|v_i\rangle$: $U |u_i\rangle = |v_i\rangle$.\end{theorem}

\textit{Proof:} $M_{ij} =\langle u_i| u_j\rangle = \langle v_i| v_j\rangle$ is a Hermitian matrix and we can diagonalize it. The possible eigenvalues are $1$ and $0$. The eigenvectors corresponding to $1$ will give a basis in $\text{span} \{ |v_i\rangle \} $ ($\text{span} \{ |u_i\rangle \} $), and the eigenvectors corresponding to $0$ will give vanishing combinations of $|v_i\rangle$  ($|u_i\rangle$). Let $c_{\alpha i}$ be the $\lambda = 1$ eigenvectors for $\alpha = 1, \dim \text{span} (\{ |v_i\rangle \})$, then the desired unitary maps $\sum_{i}c_{\alpha i} |u_{i}\rangle \to \sum_{i} c_{\alpha i} |v_{i}\rangle$ and acts trivially on the orthogonal complement of $\text{span} (\{ |u_i\rangle \})$. \textit{Q.E.D.}

With that we can prove the that principal angles characterize the configuration of two subspaces uniquely.

\begin{theorem} \label{th:pair}
Let $V, W$ and $V', W'$ be two pairs of $m$-dimensional subspaces of an $\mathbb{C}^n$ and let the respective sets of principal angles be identical: $\theta_i(V,W) = \theta_i(V',W')$ for any $i$. Then there exists a unitary transformation $g \in SU(n)$ that maps $V$ to $V'$ and $W$ to $W'$.
\end{theorem}

\textit{Proof:} 
Together with Theorem \ref{th:principal_vecs}, the hypothesis of this theorem implies 
\bea
\langle v_i|w_j \rangle = \langle v'_i|w'_j \rangle = \cos\theta_i \delta_{ij}, ~ \langle v_i|v_j \rangle = \langle v'_i|v'_j \rangle = \delta_{ij}, ~ \langle w_i|w_j \rangle = \langle w'_i|w'_j \rangle = \delta_{ij}.
\eea

Then by Theorem \ref{th:old} applied to sets $\{ v_i, w_j \}$ and $\{ v'_i, w'_j \}$, there exists a unitary $U$ such that $U |v_i\rangle = |v'_i\rangle$ and $U |w_j\rangle = |w'_j\rangle$. $U$ maps $V$ to $V'$ and $W$ to $W'$. \textit{Q.E.D.}

Similarly, we prove that the principal vectors and unitaries uniquely describe the configuration of three subspaces.

\begin{theorem} \label{th:gps_2}
Let $V, W, U$ and $V', W', U'$ be two triples of $m$-dimensional subspaces of $\mathbb{C}^n$ and let the pair-wise principal angles to be the same in both sets: $\theta(V, W) = \theta(V', W')$, $\theta(V, U) = \theta(V', U')$ and $\theta(U, W) = \theta(U', W')$. Additionally, let the principal unitaries be the same in each set: $\langle v^{W}_i | v^{U}_k \rangle = \langle v'^{W'}_i | v'^{U'}_k \rangle, ~ \langle w^V_i | w^U_k \rangle = \langle w'^{V'}_i | w'^{U'}_k \rangle, ~  \langle u^V_k | u^W_j \rangle=\langle u'^{V'}_k | u'^{W'}_j \rangle$, for some choice of principal vectors. Then there exists a unitary transformation $g \in SU(n)$ that maps $V, W$ and $U$ to $V',W'$ and $U'$.
\end{theorem}

\textit{Proof:} Our goal is find two sets of vectors that spans the span of $V, W, U$ and $V',W',U'$, respectively, and apply Theorem \ref{th:old} to them. The required sets are given by the $6 m$ principal vectors constructed in each triple of subspaces. Eq. \eqref{eq:basis_in_proof} and the unitaries give us some of the required overlaps that are equal between the sets according to the statement of the theorem and others, e.g. $\langle v^W_i | w^U_j \rangle$, can be reduced to them. The procedure is the same for all of them, and, without loss of generality, we only consider $\langle v^W_i | w^U_j \rangle$:
\bea
\langle v^W_i | w^U_j \rangle =\sum_{i'}\langle v^W_{i}| w^V_{i'} \rangle \langle w^{V}_{i'} | w^U_j \rangle = \cos\theta_i(V,W) \langle w^{V}_{i} | w^U_j \rangle,
\eea
where in the first equality we used a representation of identity in $W$ via the $|w^V_i\rangle$ basis. \textit{Q.E.D.}

And, lastly, we prove that for more than 3 subspaces we still only need principal vectors and unitaries.

\begin{theorem} \label{th:gps_3}
Let $V_l$ and $V'_l$ be two collections of $m$-dimensional subspaces in $\mathbb{C}^n$, with $1 \leq l,l' \leq L$. Furthermore, let the pair-wise principal angles and  to be the same in both sets: $\theta(V_{l}, V_{l'}) = \theta(V'_{l}, V'_{l'})$. And similarly, for the  principal unitaries (for some choice of principal vectors): $\mathcal{V}^{V_l, V_{l'}}_{i j} = \mathcal{V}^{V', V'_{l'}}_{i j}$. Then there exists a unitary transformation $g \in SU(n)$ that maps each $V_l$ to $V'_{l}$.
\end{theorem}

\textit{Proof:} Same as for Theorem \ref{th:gps_2}.

\printbibliography

\end{document}